\documentclass{emulateapj}

\newcommand{\myemail}{eduardo.gonzalez@uah.es}
\newcommand{\kms}{{\hbox {km\thinspace s$^{-1}$}}}
\newcommand{\Lsun}{{\hbox {L$_\odot$}}}
\newcommand{\Msun}{{\hbox {M$_\odot$}}}

\newcommand{\cmt}{{\hbox {cm$^{-3}$}}}
\newcommand{\cmd}{{\hbox {cm$^{-2}$}}}
\newcommand{\hdo}{{\hbox {H$_{2}$O}}}

\def\t#1#2#3#4#5#6{{\hbox {$#1_{#2#3}\!\rightarrow\!#4_{#5#6}$}}}
\def\13co{$^{13}$CO}
\def\c18o{C$^{18}$O}

\shorttitle{High-excitation OH and H$_2$O lines in Mkn 231}
\shortauthors{Gonz\'alez-Alfonso et al.}

\begin{document}


\title{High-excitation OH and H$_2$O lines in Markarian 231: the 
molecular signatures of compact far-infrared continuum 
sources\footnote{Based on 
observations with the Infrared Space Observatory, an ESA project with 
instruments funded 
by ESA Member States (especially the principal investigator countries: 
France, Germany, Netherlands, and the United Kingdom) and with the 
participation of ISAS and NASA.}}


%\author{The list}

\author{Eduardo Gonz\'alez-Alfonso}
\affil{Universidad de Alcal\'a de Henares, Departamento de F\'{\i}sica, 
Campus Universitario, E-28871 Alcal\'a de Henares, Madrid, Spain}
\email{\myemail}

\author{Howard A. Smith}
\affil{Harvard-Smithsonian Center for Astrophysics,
    60 Garden Street, Cambridge, MA 02138, USA}
\email{hsmith@cfa.harvard.edu}

\author{Matthew L. N. Ashby}
\affil{Harvard-Smithsonian Center for Astrophysics,
    60 Garden Street, Cambridge, MA 02138, USA}
\email{mashby@cfa.harvard.edu}

\author{Jacqueline Fischer}
\affil{Naval Research Laboratory, Remote Sensing Division, 
Washington, DC 20375, USA}
\email{jackie.fischer@nrl.navy.mil}

\author{Luigi Spinoglio}
\affil{Istituto di Fisica dello Spazio Interplanetario, CNR
via Fosso del Cavaliere 100, I-00133 Roma, Italy}
\email{luigi.spinoglio@ifsi-roma.inaf.it}

\and

\author{Timothy W. Grundy}
\affil{Space Science \& Technology Department, Rutherford Appleton 
Laboratory, Chilton, Didcot, Oxfordshire, OX11 0QX, UK}
\email{t.w.grundy@rl.ac.uk}


%\altaffiltext{1}{Visiting Astronomer, Harvard-Smithsonian Center for 
%Astrophysics,
%    60 Garden Street, Cambridge, MA 02138.}
%\altaffiltext{1}{Permanent address: Universidad de Alcal\'a de Henares, 
%Departamento de F\'{\i}sica, Campus Universitario, E-28871 Alcal\'a de 
%Henares, Madrid, Spain}


\begin{abstract}
The ISO/LWS far-infrared spectrum of the ultraluminous galaxy Mkn 231 
shows OH and H$_2$O lines in absorption from energy levels up to 300 K
above the ground state, and emission in the [O I] 63 $\mu$m and [C II] 158
$\mu$m lines. Our analysis shows that OH and H$_2$O are radiatively pumped by
the far-infrared continuum emission of the galaxy. The absorptions in the 
high-excitation lines require high far-infrared radiation densities, allowing 
us to constrain the properties of the underlying continuum 
source. The bulk of the far-infrared continuum arises from a warm 
($T_{{\rm dust}}=70-100$ K), optically thick ($\tau_{100\mu{\rm m}}=1-2$)
medium of effective diameter 200-400 pc. In our best-fit model of total
luminosity $L_{{\rm IR}}$, the observed OH and H$_2$O high-lying lines arise
from a  luminous ($L/L_{{\rm IR}}\sim0.56$) region with radius $\sim100$
pc. The high surface brightness of this component suggests that its infrared
emission is dominated by the AGN. The derived column densities
$N({\rm OH})\gtrsim10^{17}$ cm$^{-2}$ and $N({\rm H_2O})\gtrsim6\times10^{16}$ 
cm$^{-2}$ may indicate XDR chemistry, although significant starburst chemistry 
cannot be ruled out. The lower-lying OH, [C II] 158 $\mu$m, and [O I] 63
$\mu$m lines arise from a more extended 
($\sim350$ pc) starburst region. We show that the [C II] deficit in Mkn 231 is 
compatible with a high average abundance of C$^+$ because of an extreme 
overall luminosity to gas mass ratio. Therefore, a [C II] deficit may indicate
a significant contribution to the luminosity by an AGN, and/or by extremely
efficient star formation.
\end{abstract}




\section{Introduction}
\label{sec:intro}

The peculiar ultraluminous infrared galaxy (ULIRG, 
$L_{{\rm IR}}\ge10^{12}$ \Lsun) Markarian 231 (Mkn 231, 
12540+5708) is the most luminous infrared galaxy in the local universe, 
with a 8-1000 $\mu$m luminosity of $3.2\times10^{12}$ \Lsun\ \citep{san03}, 
and may be a representative example of the link between AGNs and 
nuclear starbursts \citep{sco04}. 
A QSO-like nucleus is evident from many observations:  optically it is 
classified as a Type 1 Seyfert \citep*{bok77,cut84,baa98},  it 
exhibits UV through IR polarization and broad absorption lines 
\citep{smi95}, it has compact X-ray emission 
\citep[e.g.,][]{gal02} and extremely compact 
mid-infrared emission \citep{soi00}, and in the radio it is 
variable and possesses a parsec scale jet \citep*{ulv99,tay99}.
Nevertheless, there is also evidence of a compact
starburst in these results as well as in 
VLA observations of H I 21 cm absorption \citep*{car98},
near-infrared observations \citep{tac02}, and
millimeter CO interferometry \citep[][hereafter DS98]{bry96,dow98}.
Estimates for the starburst luminosity 
range from 1/3 to 2/3 of the bolometric luminosity
\citep[][DS98]{dav04}. 

Molecular observations have provided important clues 
about the concentration and kinematics of the gas in Mkn 231. DS98 showed the
presence of an inner nuclear disk of radius $\sim460$ pc in CO (2-1), 
and a more extended disk with lower brightness. Most of the molecular 
gas has been found to be dense ($\sim10^4$ \cmt) and warm ($\sim70$ K)
from recent observations of CO and HCN submillimeter lines 
\citep*[][hereafter PIW07]{pap07}. 
\cite{lah07} have inferred embedded starburst chemistry in Mkn 231 and other
ULIRGs based on mid-IR Spitzer observations of ro-vibrational bands of
warm/hot HCN and C$_2$H$_2$, while \cite{gra06} and \cite{aal07} have inferred
XDR chemistry and/or radiative pumping based on 
anomalous intensity ratios of millimeter lines of HCN, HNC, and HCO$^+$.

The bulk of the luminosity in ULIRGs is emitted at far-infrared (FIR)
wavelengths, where a number of molecular tracers are detected, mostly in
absorption. Prominent lines of OH and \hdo\ 
were detected using ISO/LWS in the FIR spectrum of Arp 220, 
along with absorption features by radicals such as NH and CH,
revealing a chemistry that may be indicative of
PDRs with plausible contribution by
shocks and hot cores \citep[][hereafter Paper I]{gon04}. 
However, those species are also expected to be enhanced in XDRs
\citep{mei05}, so that the dominant chemistry in the nuclear regions of ULIRGs
remains uncertain. In Paper I, the 
ISO/LWS FIR spectrum of Arp 220 was analyzed by means of 
radiative transfer calculations, which included a non-local treatment
of the molecular excitation by absorption of FIR photons. 
Paper I showed that the population of high-excitation OH and \hdo\ 
rotational levels, in evidence from absorption in high-lying lines, 
is pumped through absorption of FIR continuum photons, a process
that requires high FIR radiation densities. The detection of these 
lines thus not only reveals the chemical and excitation conditions 
in the absorbing regions, it also sheds light on the size and 
characteristics of the underlying continuum FIR source in spite 
of the low angular resolution currently available at these wavelengths. 

In this paper we extend our approach of Paper I to the ISO/LWS 
FIR spectrum of Mkn 231, and show that this galaxy spectrum 
presents striking similarities to that of Arp 220.  Specifically,
strong absorption in the high-excitation OH and \hdo\ lines is also seen in 
Mkn 231. Rotationally excited OH in Mkn 231 has been previously detected 
via the $^2\Pi_{1/2}\,\,\Lambda$-doublet transitions \citep*{hen87}. VLBI 
observations of the mega-maser OH emission at 18 cm wavelength trace an
inner torus or disk of size $\sim100$ pc around the AGN \citep*{klo03}, 
and MERLIN observations were able to map essentially the whole single-dish 
mega-maser OH emission with angular resolution of $\approx0.3''$ 
\citep{ric05}. We analyze here both the FIR continuum emission 
and the high-excitation OH and \hdo\ lines, as well as
the [C II] 158 $\mu$m and [O I] 63 $\mu$m emission lines. 
In \S\ref{sec:obser} we present the ISO spectroscopic observations of 
Mkn 231.  In \S\ref{sec:models} we first analyze simple models for the
FIR continuum emission from Mkn 231, and then examine how well
those models reproduce the observed FIR emission and absorption
lines.  \S\ref{sec:discussion} summarizes our results. 
We adopt a distance to Mkn 231 of 170 Mpc ($H_0=75$ km s$^{-1}$ Mpc$^{-1}$ and
$z\approx0.042$).

\section{Observations and results} \label{sec:obser}

\begin{figure}
\epsscale{1}
\plotone{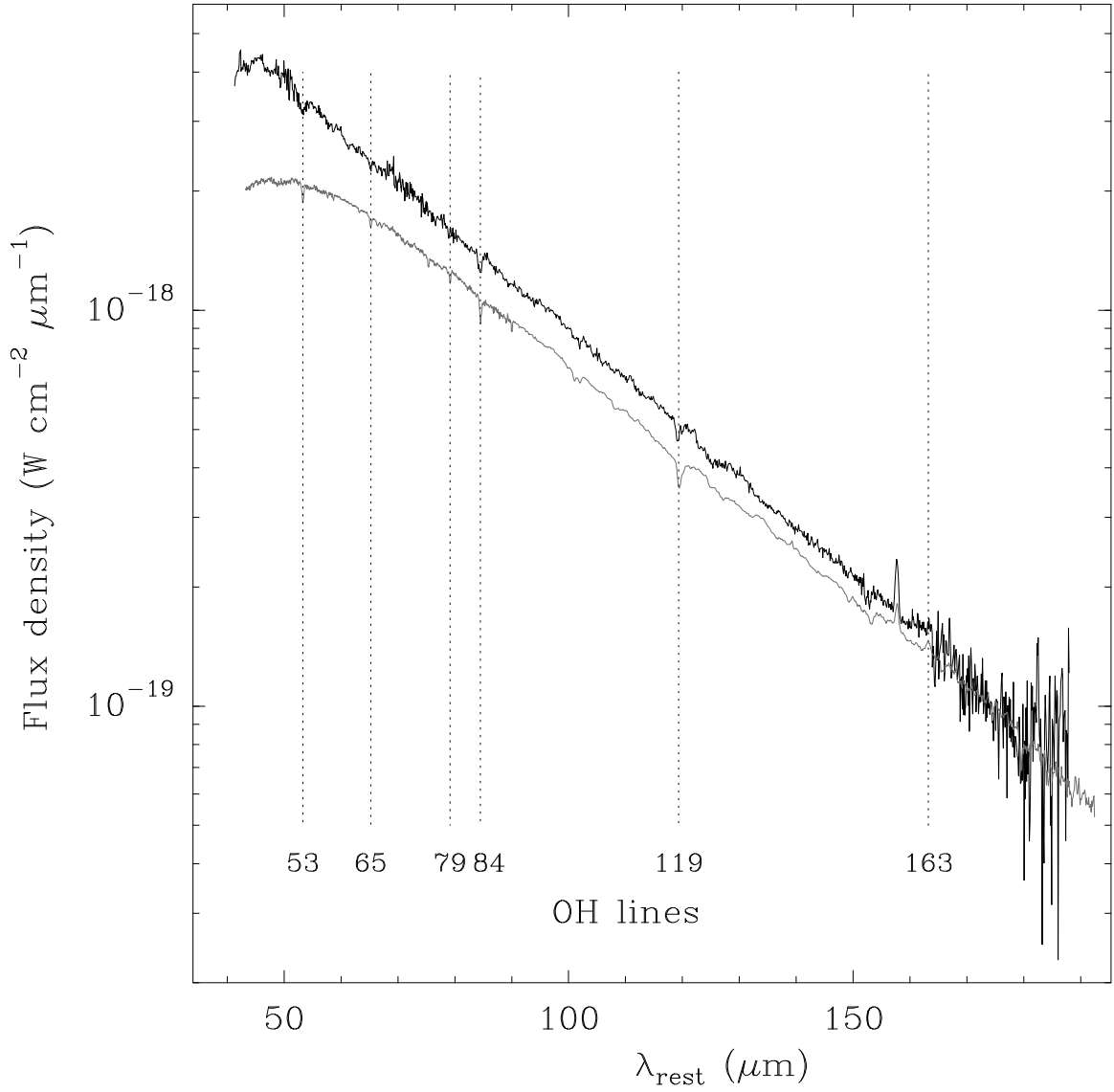}
\caption{Comparison between the FIR emission from Mkn 231 (black)
and Arp 220 (grey). The spectrum of Arp 220 has been re-scaled to the
  distance of Mkn 231 (170 Mpc).
The position of the OH lines discussed in the text are indicated and labeled
with their wavelengths. 
\label{fig:seds}}
\end{figure}

The full 43-197 $\mu$m spectrum of Mkn 231 \citep[first shown and discussed 
by][]{har99}, was obtained with the LWS spectrometer \citep{clegg96} on 
board ISO \citep{kes96}. In Fig.~\ref{fig:seds}, it is compared 
with that of Arp 220 (Paper I) re-scaled to the same distance (170 Mpc). 
The grating spectral resolution is $\sim$0.3 $\mu$m in the 43--93 
$\mu$m interval (detectors SW1--SW5), and $\sim$0.6 $\mu$m
in the 80--197 $\mu$m interval (detectors LW1--LW5), corresponding to
$\Delta v\gtrsim10^3$ \kms. The lines are thus unresolved in velocity space. 
The $\approx80''$ beam size ensures that all the FIR continuum 
and line emission/absorption from Mkn 231 (CO size $\sim4''$, DS98)
lie within the ISO/LWS aperture.

The data (TDT numbers 5100540, 18001306, and 60300241) were taken from
the highly-processed data product (HPDP) dataset (called 'Uniformly 
processed LWS01 data'), 
and reduced using version 10.1 of the Off Line Processing (OLP) 
Pipeline system \citep{swi96}. 
We performed subsequent data processing, including co-addition, scaling, 
and baseline removing, using the ISO Spectral Analysis Package 
\citep[ISAP;][]{sturm98} and our own routines. In order to obtain a smooth 
spectrum throughout the whole LWS range, the flux densities given by each 
detector were corrected by multiplicative scale factors. Corrections were 
less than 25\% except for detectors LW2 and LW3 (100--145 $\mu$m), for 
which the corrections were 30\%. We thus attribute an uncertainty
of 30\% to the overall continuum level, as well as for the line fluxes.

\begin{figure}
\epsscale{0.8}
\plotone{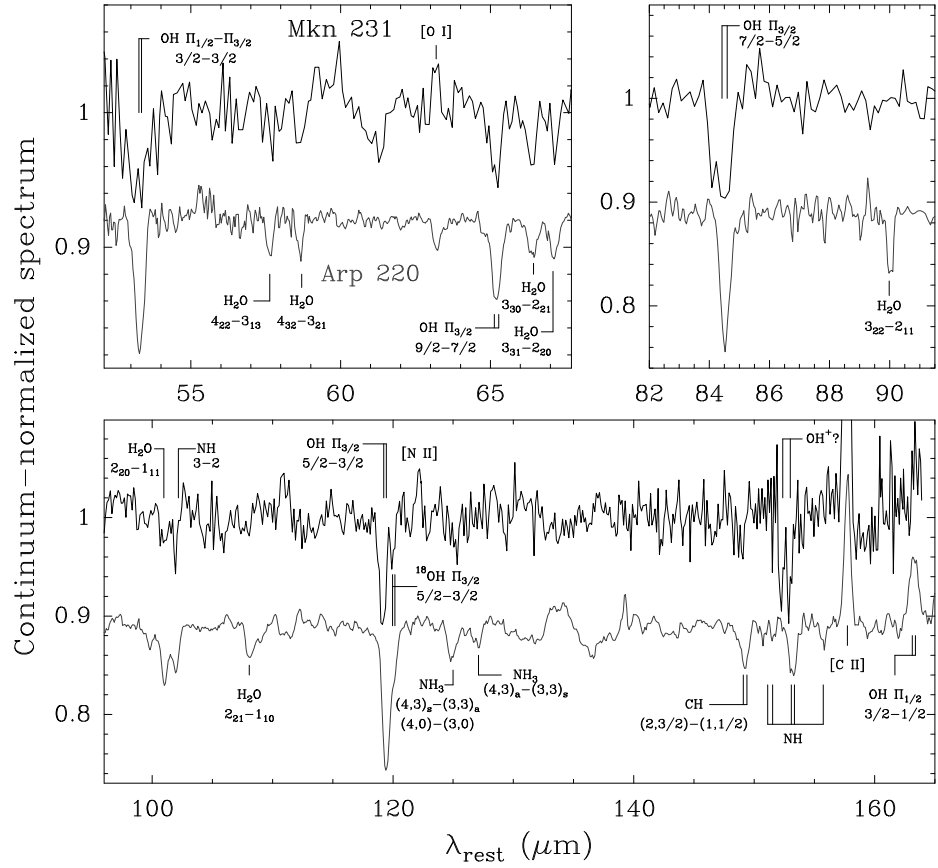}
\caption{Comparison between the continuum-normalized spectra of Mkn 231 (upper
  spectra, solid lines) and Arp 220 (lower spectra, grey lines). 
The positions of lines discussed in the text are indicated and labeled. 
\label{fig:spectra}}
\end{figure}

Figure~\ref{fig:seds} shows that the FIR spectra of Mkn 231 
and Arp 220 are similar in key aspects \citep[see also][]{fis99}, 
in particular the prominent molecular absorptions mostly 
due to OH doublets (that will be referred to hereafter as lines)
and the lack of strong fine-structure line emission typically seen 
in less luminous galaxies. A closer inspection of the pattern of line 
emission/absorption in both sources is shown in Fig.~\ref{fig:spectra}, 
where the continuum-normalized spectra are compared.
Of particular interest are the
clear detections in both sources of the high-excitation OH 
$\Pi_{3/2}\,7/2-5/2$ 84 $\mu$m and $\Pi_{3/2}\,9/2-7/2$ 65 $\mu$m lines, 
with lower level energies of 120 and 290 K, respectively 
(see \S\ref{sec:models}). The \t330221\ and 
\t331220\ \hdo\ 66-67 $\mu$m lines, both with lower levels at 195 K, are 
also detected in Mkn 231, as well as the tentatively identified
\t220111\ line at 101 $\mu$m.
It is likely that the increased noise level 
at $\lambda\gtrsim160$ $\mu$m is responsible for the non-detection of the 
high-excitation $\Pi_{1/2}\,3/2-1/2$ OH line in Mkn 231, which is seen in 
strong emission in Arp 220. 
While the high-excitation OH and \hdo\ lines at 65-67 $\mu$m are of similar
strength in Mkn 231 and Arp 220, the \hdo\ lines at longer wavelengths are
undoubtly weaker in Mkn 231, as seen for the \t322211, \t220111\ and
\t221110\ \hdo\ lines at 90, 102, and 108 $\mu$m, respectively. 
The weakness of the latter lines in Mkn 231 suggests that the region
  where the high-lying \hdo\ lines are formed is relatively weak in the far-IR
  continuum at $\lambda=90-108$ $\mu$m. The Mkn 231
spectrum thus suggests that a warm component, with relatively weak 
contribution to the far-IR continuum at 
$\lambda\gtrsim80$ $\mu$m, is responsible for the observed high-excitation
absorptions (\S\ref{sec:molec}). 
Table~\ref{tab:flux} lists the line fluxes, continuum flux densities at the
corresponding wavelengths, and equivalent widths for the lines detected in Mkn
231.

In the case of Arp 220, we used high-spatial resolution continuum measurements 
available in the literature to infer that Arp 220 is optically 
thick even in the submillimeter continuum \citep[Paper I; see also][]{dow07}.
The steeper decrease of the flux density with increasing wavelength
in Mkn 231, however, suggests that it has lower FIR continuum 
opacities (Fig.~\ref{fig:seds}). This expectation is further reinforced
by the detection in Mkn 231 of the [N II] 122 $\mu$m line, a feature not
seen in Arp 220 (Fig.~\ref{fig:spectra}). 
Other notable differences between both sources are
that the [O I] 63 $\mu$m line is observed in emission in Mkn 231 but in
absorption in Arp 220, and that the ground-state 119, 53, and 79 $\mu$m OH 
lines are significantly weaker in Mkn 231 (Fig.~\ref{fig:spectra}). In 
modeling Arp 220, we were forced to invoke an absorbing ``halo'' to account 
for these lines; in Mkn 231, no such halo is required 
(\S\ref{sec:models}).

In the spectrum of Mkn 231,
the main 119.3 $\mu$m OH line appears to be slightly blue-shifted relative to
the expected position, an effect we attribute to the proximity of the
line to the edge of the LW3 detector. There is a nearby weaker red-shifted 
feature, at 120 $\mu$m, which coincides
with the expected position of the ground $\Pi_{3/2}\,5/2-3/2$ $^{18}$OH
line, and appears as a marginal feature in both the ``up'' and 
``down'' grating scans. However, the limited signal-to-noise ratio 
($(1.0\pm0.4)\times10^{-20}$ W cm$^{-2}$), the narrow appearance of the 
feature ($\approx0.42$ $\mu$m), and
the fact that it is not blue-shifted as the main line, make that assignment
only tentative. In Arp 220, the main OH line is not shifted because it does 
not fall so close to the edge of the detector, as a consequence of the 
lower red-shift of the source. In Arp 220, a red-shifted shoulder 
appears at 120 $\mu$m, suggesting the 
possibility that $^{18}$OH may be responsible for it (Paper I). 
We cannot however be certain that $^{18}$OH is detected in any of these 
sources, but given the high $^{16}$OH column densities we derive in some
  of our models below (\S\ref{sec:molec}) and the fact that values of the
  $^{16}$OH/$^{18}$OH ratios below the canonical value of 500 may be expected
  in regions where the ISM is highly processed by starbursts (Paper I), our
  tentative identification should be followed up with future Herschel Space
  Observatory observations with higher spectral resolution and sensitivity.
Finally, the spectrum of Mkn 231 shows a broad feature at the 
position of the $\Pi_{1/2}-\Pi_{3/2}\,3/2-3/2$ OH line (53 $\mu$m). We 
note that the blue-shifted side of this absorption is coincident with the OH 
$\Pi_{3/2}\,11/2-9/2$ line, with a lower level energy of 511 K; however, the 
proximity of this spectral feature to the edge of the SW2 detector 
precludes any definitive assignment.

The FIR detections of both NH and NH$_3$ in Arp 220 were reported in
Paper I. NH$_3$ was also detected via 
the 25 GHz inversion transitions by \cite{tak05}, who derived
a NH$_3$ column density six times higher than our value.  
The difference likely arises because of the high FIR 
continuum opacities in Arp 220, which cause the observed FIR 
absorptions to trace only a fraction of the total gas column. Since there are 
no such extinction effects at 25 GHz, the NH$_3$ inversion transitions are 
expected to trace higher NH$_3$ column densities.
Figure~\ref{fig:spectra} shows that, by contrast, the
NH$_3$ lines are not detected in Mkn 231, although the relatively 
high noise at 125 $\mu$m
does not rule out future detection of NH$_3$ with Herschel at a
level similar to that of Arp 220. 

\begin{figure}
\epsscale{0.8}
\plotone{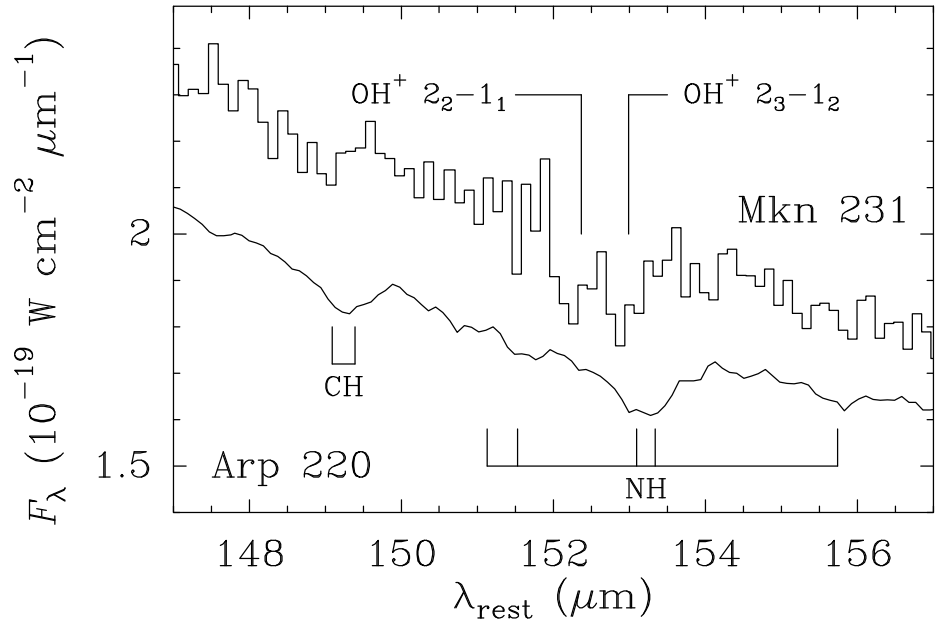}
\caption{Comparison between the spectra of Mkn 231 (upper
  spectrum, histogram) and Arp 220 (lower spectrum, solid line)  
  around 152 $\mu$m. The spectrum of Arp 220 has been re-scaled to the
  distance of Mkn 231 (170 Mpc).
  The position of the NH, CH, and OH$^+$ lines 
  are indicated and labeled. 
\label{fig:ohplus}}
\end{figure}

There are two marginally-detected ($2.5\sigma$ level) spectral features 
seen at 153.0 and 152.3 $\mu$m, in the Mkn 231 spectrum 
(Fig.~\ref{fig:ohplus}).
Although close to the expected position of the main NH feature at 153.22 
$\mu$m, the 153.0 $\mu$m feature appears significantly shifted by 0.25 
$\mu$m from it, and better coincides with the position of the 
OH$^+$ $2_3-1_2$ line. Also, the 152.3 $\mu$m feature lies at 0.1 $\mu$m from
the expected position of the OH$^+$ $2_2-1_1$ line. In Paper I, we also
suggested that OH$^+$ could contribute to the spectrum of Arp 220 for two
reasons: $(i)$ our models were unable to 
reproduce, using NH and NH$_3$, the observed strong absorption at 
102 $\mu$m, which coincides with the expected position of the
OH$^+$ $3_4-2_3$ line; $(ii)$ there was an absorption feature at 76.4 
$\mu$m that, if real, could be attributed to the OH$^+$ $4_4-3_3$
transition. 
Since OH$^+$ has never been detected in the galactic interstellar medium 
or that of any galaxy, here we only highlight the intriguing possibility 
of its detection in two ULIRGs. Sensitive, higher-resolution Herschel 
observations are needed to resolve this tantalizing speculation.

The luminosity of the [C II] $^2P_{3/2}-^2P_{1/2}$ fine-structure line 
at 158 $\mu$m is 2.5 times stronger in Mkn 231 than in Arp 220, but given 
the higher FIR luminosity of this source (Fig~\ref{fig:seds}), 
the [C II] to FIR luminosity ratios are rather similar, with 
values of $2.5\times10^{-4}$ and 
$2.1\times10^{-4}$ for Mkn 231 and Arp 220, 
respectively \citep{luh03}. These are among the lowest values found
in galaxies, illustrating the so-called ``[C II] deficit'' found in
ULIRGs. The [C II] line emission from Mkn 231
is analyzed in \S\ref{sec:cii}.

\section{Analysis} \label{sec:models}

\subsection{Models for the far-infrared continuum} \label{sec:continuum}

\begin{figure}
\epsscale{0.8}
\plotone{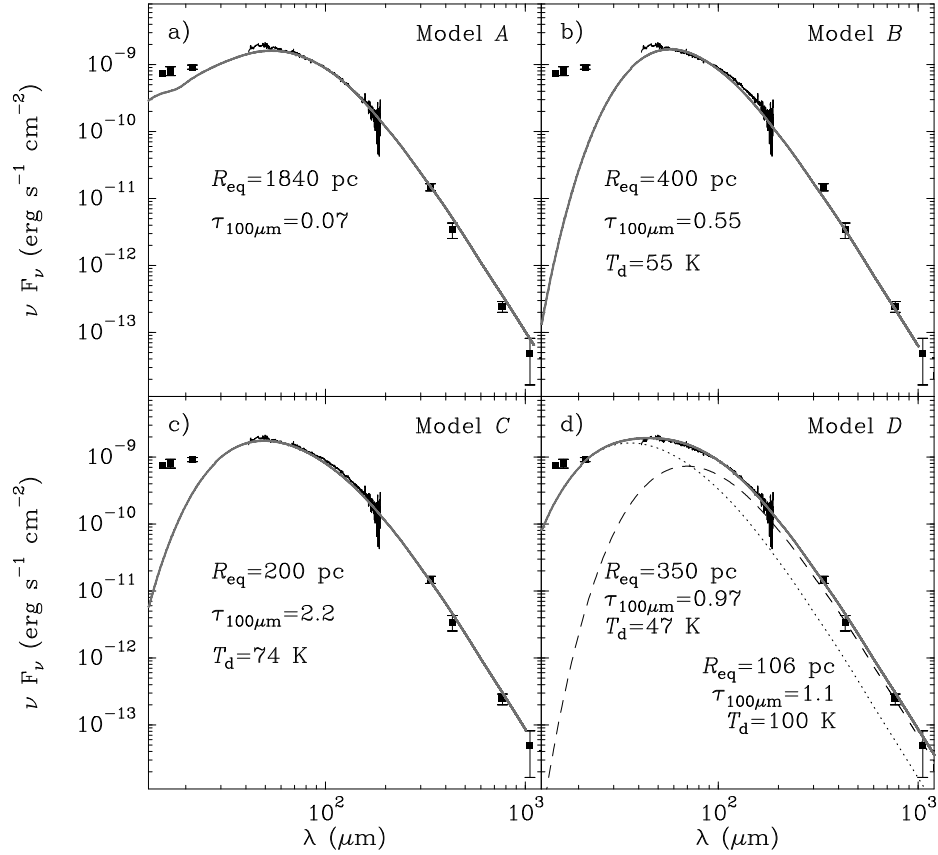}
\caption{Different models for the FIR continuum emission
from Mkn 231. The whole set of parameters that characterize the models $A$,
$B$, $C$, and $D$, are listed in Table~\ref{tab:cont}. Flux densities at 800
and 1100 $\mu$m are taken from \cite[][corrected for non-thermal
emission]{roc93}, at 450 $\mu$m from \cite*{rig96}, at 350 $\mu$m
from \cite{yan07}, and at $\lambda<25$ $\mu$m from \cite{rie76}.
\label{fig:cont}}
\end{figure}

Figure~\ref{fig:cont} illustrates several ways that the FIR to 
millimeter continuum can be fit and interpreted.
We first modeled (model $A$ in Fig.~\ref{fig:cont}a) the far infrared 
source in Mkn 231 as an ensemble of identical dust clouds each of 
which is heated by its own single 
central luminosity source. The representative cloud is assumed to be 
spherical, with radius $R_c$, and is divided into concentric shells whose dust 
temperatures are computed from the balance of heating and cooling 
\citep{gon99}. We used a mixture of silicate and amorphous carbon grains 
with optical constants from \cite{pre93} and \cite{dra85}.
The stellar continuum was taken from \cite{lei99}, but results 
depend only weakly on this choice because the intrinsic continuum is  
absorbed by the dust and re-emitted at infrared wavelengths. Once the 
equilibrium temperatures are obtained for each shell, the resulting continuum 
emission from the cloud is computed, and multiplied by $N_c$, the 
number of clouds in the source required to match the absolute flux densities. 
This scaled spectrum is shown in Fig.~\ref{fig:cont}a. 
The other three models ($B$, $C$, and $D$, shown in
Fig.~\ref{fig:cont}b-d)  use grey-bodies with uniform dust temperatures
$T_d$ to characterize the continuum emission \citep[e.g.,][]{roc93,arm07}.

Assuming that the individual clouds do not overlap along the line of
sight, our results do not depend particularly on the radius or 
luminosity adopted for the model individual cloud because identical results
are obtained if $R_c$ is multiplied by a factor of $\alpha$, the luminosity by
$\alpha^2$, $N_c$ by $\alpha^{-2}$, and the continuum opacity is kept constant
(see Paper I). The models are thus characterized by the luminosity of the
whole ensemble, the radial opacity of the clouds at a given wavelength (which 
we adopt to be 100 $\mu$m: $\tau_{100\mu{\rm m}}$), and the equivalent 
radius of the source, defined as $R_{eq}=N_c^{1/2} R_c$. These parameters are
listed in Table~\ref{tab:cont}. 
 
In model A, the individual clouds are optically thin so that some degree
of cloud overlap would yield a similar fit to the continuum while decreasing
the value of $R_{eq}$. For instance, if the clouds are
distributed in a spherical volume, $R_{eq}=N_c^{1/3} R_c$ giving
$R_{eq}=400$ pc for clouds with $R_c=20$ pc. However, the predicted opacity
through the modeled region, $N_c^{1/3}\tau_{100\mu{\rm m}}$, will be much
higher than that of an individual cloud, and this physical situation
is already described in models $B$-$C$ where higher opacities along the line
of sight and a more compact region of FIR emission are assumed. In order to
avoid this model redundancy, we 
choose our continuum models such that an individual ``cloud''
describes the characteristic continuum opacity ($\tau_{100\mu{\rm m}}$ in
Table~\ref{tab:cont}) and dust temperature through the whole region (disk),
so that the resulting extent of the FIR emission is $R_{eq}=N_c^{1/2} R_c$.

The observed continuum can be reproduced from model $A$'s cloud ensemble 
that is optically thin in the FIR. Model $A$ also predicts that the starburst
dominates the continuum for $\lambda\gtrsim15$ $\mu$m, while 
the torus/disk around the AGN would then dominate the mid-infrared continuum,
in qualitative agreement with the models by \cite{far03}.
The equivalent radius of the starburst is slightly larger than the radius of 
the outer disk observed by DS98. Because $\tau_{100\mu{\rm m}}$ is low and 
$R_{eq}$ is high, this model predicts that the 
{\em FIR radiation density} is low, a prediction that is not 
consistent with our models of the observed OH line strengths 
(\S\ref{sec:eqw}).

As both $\tau_{100\mu{\rm m}}$ and $T_d$ are increased in models $B$ and $C$,
the radiation density increases and, therefore, the
equivalent size required to reproduce the observed emission becomes smaller.
As a consequence, models $B$ and $C$ predict increasing compactness of the
dust clouds responsible for the FIR emission, with $R_{eq}=400$ 
and 200 pc respectively. With a single-component model, however, 
$R_{eq}$ cannot be reduced more than in model $C$ without degrading the 
quality of the fit. However, a two-component model as shown in $D$ 
is able to reproduce the FIR emission, invoking a quite 
compact ($\sim100$ pc) and warm (100 K) component ($D_{{\rm warm}}$), and 
a colder and more extended one that dominates at $\lambda>80$ $\mu$m
($D_{{\rm cold}}$).

A convenient way to characterize the radiation density in the modeled
regions is to compute the radiation temperature at 100 $\mu$m from
\begin{equation}
T_{{\rm rad}}(100\,\mu{\rm m})=
\frac{h\nu}{k\,\ln\left[1+\frac{2h\nu^3\Omega}{c^2F_{100\,\mu{\rm m}}}
  \right]},
\end{equation}
where $\Omega=\pi N_c R_c^2/D^2$ is the solid angle subtended by the modeled 
source, $F_{100\,\mu{\rm m}}$ is the predicted flux density at 100 $\mu$m, 
and other symbols have their usual meaning.  $T_{{\rm rad}}(100\,\mu{\rm m})$ 
is also listed in Table~\ref{tab:cont}, together with the gas mass, 
luminosity, and fraction of the bolometric luminosity for each model. 
The calculated gas masses assume a gas-to-dust mass ratio of 100. 
In all cases, they are lower than the dynamical masses determined by DS98 
when $R_{eq}$ is identified with the radial 
extent of the source (and therefore compatible with the inferred 
rotation velocities in the disk). Our inferred masses are in models $B-D$
consistent with the mass inferred by PIW07, but are in all cases
higher, by at least a factor of two, than the gas masses 
obtained by DS98. This discrepancy may be explained in at least four possible, 
different ways: $(i)$ the physical radial extent
of the cloud ensemble, which accounts for cloud filling,
is given by $R_T=f^{-1/2}R_{eq}$, where $f$ is the area filling 
factor, so that $R_{eq}$ is a lower limit of $R_T$; $(ii)$ our 
calculated masses depend on the mass-absorption coefficient for dust, which 
we have assumed to be $\kappa_{1300\mu{\rm m}}=0.33$ cm$^2$ g$^{-1}$ based
on a mixture of silicate and amorphous carbon grains \citep{pre93,dra85}, but 
could be up to a factor $\sim6$ higher if the dust is mainly composed 
of fluffy aggregates \citep{kru94}; $(iii)$ the gas-to-dust mass ratio 
may depart significantly from the standard value of 100; $(iv)$ the masses 
derived by DS98 for Mkn 231 could be lower limits in the light of the 
submillimeter CO emission reported by PIW07. A combination of these factors
may explain our higher values.

The luminosities in Table~\ref{tab:cont} account for 50-80 \% of the observed
$8-1000$ $\mu$m infrared luminosity. Model $A$ implicitly assumes that the
calculated luminosity has a starburst origin; the luminosity from 
model $B$ and from the cold component of model $D$ are also attributable 
to the starburst in view of the spatial extent of the modeled source. 
Since model $C$ and the warm component of model $D$ are more compact, a
combination of AGN and starburst contributions is more plausible. 
The surface brightness in model $C$ is $4\times10^{12}$ \Lsun/kpc$^2$, 
a factor of 2 higher than the peak global value found
in starburst galaxies by \cite{meu97}, suggesting an important (but uncertain) 
contribution by the AGN to the observed FIR emission \citep{soi00}. 
Also, the luminosity-to-mass ratio of 500 \Lsun/\Msun\ coincides 
with the uppermost limit proposed by \cite{sco04} for a starburst. 
The very high surface brightness ($1.3\times10^{13}$ \Lsun/kpc$^2$) and 
luminosity-to-mass ratio ($\sim3300$ \Lsun/\Msun) of the warm component of 
model $D$ ($D_{{\rm warm}}$), as well as its compactness, persuasively 
indicate that this component is most probably dominated by the AGN. 
The most plausible relative contributions by the AGN and the starburst 
to $D_{{\rm warm}}$ are discussed in \S\ref{sec:discussion}.

In summary, different approaches can be used to successfully fit the observed
FIR continuum emission, with the properties of the clouds that
emit that radiation in these approaches spanning a wide range of possible
physical scenarios. But ISO/LWS has provided us with spectroscopic information,
and we show next how the observed high excitation OH and \hdo\ 
lines impose important constraints on these continuum models.

\subsection{Equivalent widths} \label{sec:eqw}

We analyze the OH equivalent widths assuming that the OH molecules 
form a screen in front of the IR source. The strengths of the 
$\Pi_{3/2}$ $7/2-5/2$ and $9/2-7/2$ OH doublets at 84 and 65 $\mu$m, 
enable us to conclude that the excited OH covers a 
substantial fraction of the FIR emission region. 
Assuming that each line of the 84 $\mu$m doublet absorbs all the background 
84 $\mu$m continuum over a velocity range of 250 km s$^{-1}$ 
{\em along each line of sight}, and that there is no significant re-emission in
the line, the covering factor is $\sim50$\%. This value may be considered
a lower limit for the following reasons. The submillimeter CO line 
profiles shown by \cite{pap07} have FWHMs of 200-250 km s$^{-1}$, and the 
lines are expected to be broadened by velocity gradients and, in particular, 
by the disk rotation; therefore, the velocity range of 250 km s$^{-1}$ assumed
above is probably an upper limit. DS98 inferred local turbulent velocities 
of up to 60 km s$^{-1}$ at inner radii (100 pc) and decreasing as $r^{-0.3}$. 
If we adopt an intrinsic Gaussian line profile with
the highest value of the turbulent velocity, 
$\Delta V=60$ km s$^{-1}$, and saturate the 84 $\mu$m line to the degree
that an effective width\footnote{The effective width is defined here as
$\int (1-\exp\{-\tau_v\})dv$, where
$\tau_v=\tau_0\times\exp\{-(v/\Delta V)^2\}$ and $\tau_0$ is the line opacity
at line center.} of 250 km s$^{-1}$ is obtained for each component of
the doublet, the derived 84 $\mu$m foreground opacity at line center is
$\sim50$, but the high column density required for this opacity is 
hard to reconcile with that inferred from the
other observed OH line strengths (\S\ref{sec:molec}). Finally, 
some significant re-emission in the 84 $\mu$m OH line is expected because the  
$\Pi_{3/2}$ $9/2-7/2$ OH line at 65 $\mu$m that originates from its upper
level is detected in absorption. We therefore conclude
that the observed 84 $\mu$m OH absorption is widespread, and probably covers
the bulk of the 84 $\mu$m continuum emission regions. On the other hand, the 
opacities in the high-lying 65 $\mu$m line should only be moderate; 
for reference, if we adopt for each component an upper limit of 150 km
s$^{-1}$ on the effective velocity interval for the absorption at each sight
line, the minimum covering factor for this line is then 25\%. It is therefore
possible that the OH responsible for the 65 $\mu$m absorption does not entirely
coincide with that producing the 84 $\mu$m absorption but is only a
fraction of the latter, consistent with its lower energy level being at
nearly 300 K. Nevertheless, for the sake of simplicity, we 
assume in this Section that both lines arise in the same region --one that, 
on the basis of the 84 $\mu$m OH strength, covers the total FIR 
continuum. The derived OH column densities will be lower limits, and the
inferred properties of the continuum source will be associated with at least 
$\sim50$ \% of the observed FIR emission.

The equivalent widths $W$ are then given by
\begin{equation}
W=2\times \left[ 1-\frac{B_{\nu}(T_{ex})\Omega}{F_{\lambda}} \right]
\times \int (1-\exp\{-\tau_v\})\, dv,
\label{eq:w}
\end{equation}
where $B_{\nu}(T_{ex})$ is the blackbody emission at the excitation
temperature $T_{ex}$ of the line, $\Omega=\pi R_{eq}^2/D^2$ is the solid angle
subtended by the source, $F_{\lambda}$ is the observed continuum flux density
at the wavelength $\lambda$ of the line, $\tau_v$ is the line opacity at
velocity $v$, and the factor 2 accounts for the two lines that compose a
doublet. The values of $W$ are positive for absorption lines, and negative for
lines observed in emission.
Equation (\ref{eq:w}) applies both to optically thin and 
optically thick lines. For optically thin lines, $W$ is proportional to the
assumed column density; for very optically thick lines, 
$W$ becomes insensitive to 
the column density and scales linearly with the turbulent 
velocity.  Based on DS98, we have adopted $\Delta V=40$ km s$^{-1}$, 
which is probably accurate within 50\%.

The fractional level populations and column densities 
(and hence the opacities), and 
blackbody temperatures required to obtain $W$ from eq.~(\ref{eq:w}) 
can be estimated from a given column density $N$(OH), and 
by assuming that {\em the excitation temperature $T_{ex}$ 
is the same for all OH transitions}. This assumption is certainly only an
approximation. The OH levels are pumped primarily 
through absorption of FIR photons; this is quite a general model 
result when both the ground and the excited OH rotational lines (except the 
163 $\mu$m one) are observed in absorption \citep[see the similar conclusion 
in Paper I for the case of Arp 220, as well as the case of the \hdo\ lines 
at 25-45 $\mu$m in Orion/IRc2,][]{wri00}.
When radiative excitation dominates, the level populations tend toward 
equilibrium with the radiation field within the inner cloud, and the
excitation temperatures are similar for all transitions in those
regions. However, the observed absorptions are produced close to the 
cloud boundary where the OH molecules are illuminated from only one side and
$T_{ex}$ is less than the radiation temperature. Due to trapping
effects, the lines with the highest opacity (119 and 84 $\mu$m OH lines) 
remain more excited than the other lines at inner locations of the region, but 
their $T_{ex}$ decrease more steeply outwards and fall below the $T_{ex}$ of
other, thinner lines close to the cloud boundary. The detailed, non-local
radiative transfer models described in 
\S\ref{sec:molec} show that the assumption of
equal $T_{ex}$ is only approximately valid for the 119 and 84 $\mu$m 
lines, but for the other lines, and in particular for the 65 and 53 $\mu$m
lines, $T_{ex}$ may be higher or lower depending on position in the region. 
Nevertheless, the opacities of the 119, 84, 65, and 53 $\mu$m lines
are mostly determined by the $T_{ex}$ of the 119 and 84 $\mu$m lines. As a
result, the following analysis of the radiation 
density (or, equivalently, $R_{eq}$, required to account for the observed 
absorptions) is accurate, at least to a first approximation.

\begin{figure}
\epsscale{1}
\plotone{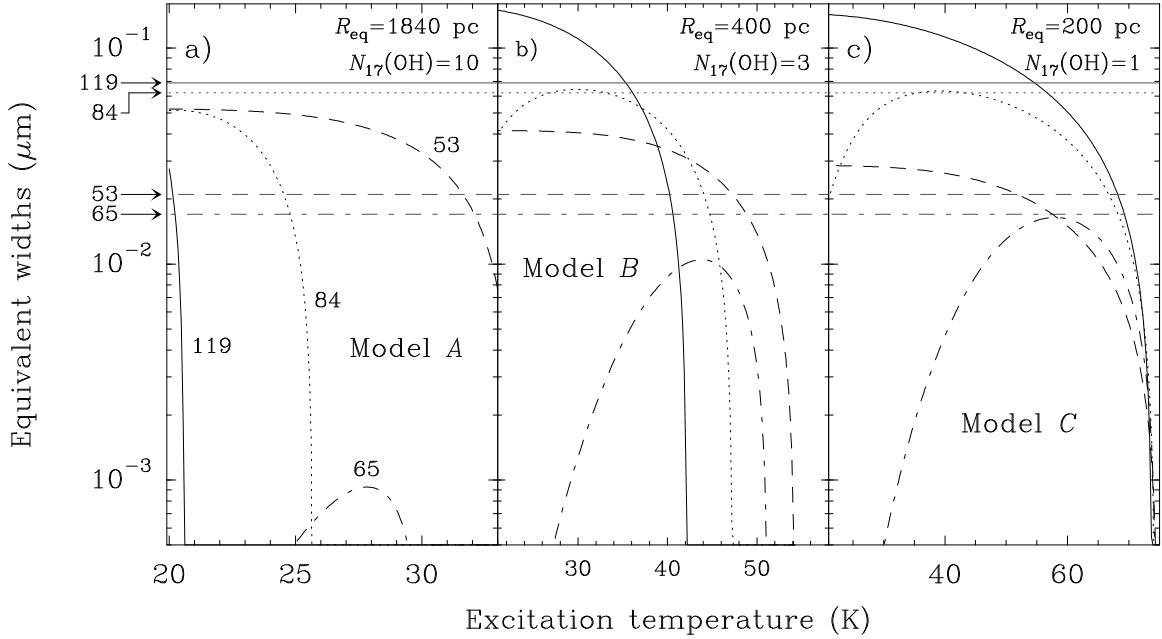}
\caption{Predicted equivalent widths ($W$) versus the OH excitation temperature
  ($T_{ex}$) for the 119 (solid curves), 
  84 (dotted), 65 (dashed-dotted), and 53 (dashed) $\mu$m OH doublets. Results
  are shown for three assumed sizes of the continuum emission, corresponding 
  to the continuum models $A$, $B$, and $C$ in Table~\ref{tab:cont}. 
  The observed values are indicated with
  horizontal lines. $N_{17}({\rm OH})$ is the assumed OH column density in
  units of $10^{17}$ cm$^{-2}$. The equivalent widths are positive if the
  lines are in absorption; a drop in $W$ with increasing $T_{ex}$ thus 
  indicates that the corresponding line is turning from absorption to emission.
\label{fig:eqw}}
\end{figure}

The three parameters now required to estimate $W$ from equation (\ref{eq:w}) 
are then $N$(OH), $R_{eq}$, and $T_{ex}$. Figure~\ref{fig:eqw} shows the 
expected values of $W$ for models $A$, $B$, and $C$ as a function
of $T_{ex}$ for the 119, 84, 65, and 53 $\mu$m lines, and compares them
with the observed values. Each panel assumes a value of $R_{eq}$ that
corresponds to the continuum models $A$, $B$, and $C$ described in
\S\ref{sec:continuum} (see also Table~\ref{tab:cont}). The values of
$N$(OH) used for each model are just reference values discussed below.

The spectral line analysis of Model $A$ (Fig.~\ref{fig:eqw}a) shows that 
this scenario can be ruled out as the main
source of FIR radiation from Mkn 231 (Fig.~\ref{fig:cont}a): 
the continuum model predicts low $T_{rad}$ (between 20 and 34 K for the 
different lines) but the observed absorption in the 65 $\mu$m line, with 
lower level energy of 290 K, requires a much higher $T_{ex}$. 
Furthermore, the 119 and 84 $\mu$m lines are expected to
be in emission ($W<0$) as soon as $T_{ex}$ becomes higher than 20 and 25 K
(which, on the other hand, is not possible for radiative excitation).
With such a low value of $T_{ex}$, $W_{65\mu{\rm m}}$ is negligible even 
with our adopted screen $N({\rm OH})=10^{18}$ cm$^{-2}$, a value that
overestimates $W_{53\mu{\rm m}}$ by more than a factor of two.

The above problems still remain to some extent in model $B$, 
when $R_{eq}$ is reduced to 400 pc (Fig.~\ref{fig:eqw}b). Here 
the dust radiation temperatures allow the lines to
be seen in absorption up to $T_{ex}\approx40$ K, however $W_{65\mu{\rm m}}$ is
underestimated by more than a factor of 2 for the adopted 
$N({\rm OH})=3\times10^{17}$ cm$^{-2}$, yet this column density still 
overestimates the absorption of the 53 $\mu$m line. Although model $B$ cannot 
account for the 65 $\mu$m line strength, a region of similar size but 
lower $N({\rm OH})$ could contribute to the observed absorptions of the 
119, 84 and 53 $\mu$m lines.

The single-component model that best accounts for the four observed 
OH lines is model $C$ with $R_{eq}=200$ pc (Fig.~\ref{fig:eqw}c). 
The corresponding continuum model (Fig.~\ref{fig:cont}c), with $T_d=74$ K, 
also fits rather well the overall FIR continuum emission. 
Significantly, our models in \S\ref{sec:molec} show that 
the excitation temperatures required to
reproduce the observed equivalent widths, 40-60 K, are those computed at the
cloud surface if the OH is excited by the infrared emission from a blackbody 
at $T_d=74$ K. Finally, the dust temperature and gas mass 
(Table~\ref{tab:cont}) in model $C$ are consistent with the gas 
temperature and H$_2$ mass derived 
by \cite{pap07} from the submillimeter CO and HCN emission. They 
found that this warm gas component hosts most of the molecular mass in 
the galaxy. The H$_2$ column density, $N({\rm H_2})\sim1.5\times10^{24}$ 
cm$^{-2}$, indicates high optical depths, as in the galactic Sgr B2 
molecular cloud, but Mkn 231 is much warmer. If the column density in Mkn 231
is concentrated in a face-on disk of thickness $H=23$ pc, 
as concluded by DS98, the expected density is $n({\rm H_2})\sim2\times10^4$ 
cm$^{-3}$, just the amount needed to account for the CO submillimeter lines 
\citep{pap07}. On the other hand, if this warm and dense component is 
identified with the inner disk of radius 460 pc reported by DS98, the 
area filling factor is $f\sim0.2$. In spite of the general agreement between
our model $C$ with other observations, a closer 
inspection of this model (\S\ref{sec:molec}) reveals some 
discrepancies with other OH and \hdo\ lines that suggest that a slightly 
modified scenario can better explain the overall observed absorption 
patterns.

\subsection{Models for OH and \hdo} \label{sec:molec}

Radiative transfer modeling of the observed OH and \hdo\ lines was done 
using the code described in \cite{gon97,gon99}, which computes the 
statistical equilibrium populations of a given molecule in spherical 
symmetry. Line broadening is assumed to be caused by microturbulence. Our code 
accounts for a non-local treatment of the radiative trapping in the 
molecular lines and of the excitation through absorption of photons emitted
by dust, as well as for collisional excitation. 
Both line and continuum opacities for photons emitted in both 
lines and continuum are taken into account. 
Collisional rates were taken from \citet*{offer94} and 
\citet*{green93} for OH and \hdo, respectively. As we also found for
Arp 220 (Paper I), the overall excitation is dominated by absorption of 
FIR continuum photons in all models. If shock conditions (high 
density and temperature) were assumed, only the absorption in the 
lowest-lying lines would be significantly affected. Once the continuum model is
fixed, our results only depend on the molecular column densities and 
turbulent velocity (see Paper I for a fuller description). 

As mentioned above (\S\ref{sec:eqw}), the observed absorption strengths
are not sensitive to the amounts of OH and \hdo\ in the inner regions of 
the modeled regions, but only to the amounts of OH and \hdo\ that are 
close to the cloud (or disk) boundary. For this reason, we calculate two 
values for the derived molecular column densities: $N^{{\rm scr}}(X)$ 
denotes the column density for a shell of species $X$ covering the infrared 
source (i.e., the screen case), whereas $N^{{\rm mix}}(X)$ is the inferred 
column density for models where $X$ and dust are evenly mixed (the mixed 
case). Evidently $N^{{\rm mix}}$ will be much higher than $N^{{\rm scr}}$, 
but from our data there are only a few, non-definitive ways to discriminate 
between the alternatives. The 163 $\mu$m OH and 120 $\mu$m $^{18}$OH lines are
stronger in the mixed case, but neither of these features is unambiguously
detected. Nevertheless, we do not find any strong arguments for thinking that
OH and \hdo\ are only present on the surface of the disk, and so the 
$N^{{\rm mix}}$ values may be considered somewhat more reliable. The 
abundances we derive below are based on this assumption; we revisit the 
``mixed''case when we discuss models for the [C II] line.

Since model $C$ (Fig.~\ref{fig:cont}c, Fig.~\ref{fig:eqw}c,
Table~\ref{tab:cont}) gives the best single-component 
fit to most of the OH equivalent widths,
we first check if it can account for the observed OH and \hdo\ absorption 
features. Figure~\ref{fig:model} compares the observed continuum-subtracted 
spectrum and the modeled results (dashed spectrum, mixed case) for the 
wavelength ranges where the signal-to-noise ratio is adequate. 
Table~\ref{tab:model} lists the physical parameters obtained for this 
model. We have assumed a turbulent velocity $\Delta V$ of 40 km s$^{-1}$ 
(\S\ref{sec:eqw}; DS98). The model fits satisfactorily 
the OH 119, 84, 65, and 53 $\mu$m lines, thus demonstrating the 
approximate validity
of the simple method outlined in \S\ref{sec:eqw}. The value of
$N^{{\rm scr}}({\rm OH})=10^{17}$ cm$^{-2}$ is also the same as estimated
from the equivalent widths. The \hdo\ column densities are determined by 
the strengths of the \t330221\ and \t331220\ lines at 66-67 $\mu$m.

Some features of model $C$, however, are inconsistent 
with the data. The possible emission in the OH 163 $\mu$m line is not
reproduced, and the absorption in the 79 and 99 $\mu$m OH lines appears
excessive. The model also predicts too much absorption in the \t322211\ 
(90 $\mu$m), \t220111\ (101 $\mu$m), \t221110\ (108 $\mu$m), and \t414303\ 
(113 $\mu$m) \hdo\ lines. All these discrepancies suggest that the component 
that accounts for the absorption of the 65-68 $\mu$m OH and \hdo\ lines
is weaker than postulated in model $C$ at wavelengths longer than 80 $\mu$m.

These discrepancies may be resolved by invoking two different
components for the FIR continuum emission, as in model $D$ 
(Fig.~\ref{fig:cont}, Table~\ref{tab:cont}). The 
warm-compact component, responsible for the 65-68 $\mu$m OH and 
\hdo\ lines, will produce weak absorptions in the 80-120 $\mu$m range
as a consequence of the relatively weak continuum emission at these
wavelengths. The more extended component will contribute to the
observed absorptions in the 53, 84, and 119 $\mu$m OH lines. The compactness
of the warm component suggests that it is relatively close to the AGN, and
thus we have assumed $\Delta V=60$ \kms\ for $D_{{\rm warm}}$ 
(Table~\ref{tab:model}); this is the turbulent velocity found by DS98 
around the rotation curve turnover radius 
of 75 pc. For the extended component ($D_{{\rm cold}}$), 
$\Delta V=40$ \kms\ is assumed. 
Figure~\ref{fig:model} shows that a better fit to the overall spectrum is
indeed found with this composite model (grey line), with the lines in 
the 80-120 $\mu$m range brought down to levels compatible with observations. 
Also, the model predicts the 163 $\mu$m OH line to be in emission. The column 
densities in model $D$ are significantly lower than in $C$ and, to avoid 
the above mentioned \hdo\ absorptions at 90-120 $\mu$m, only an upper 
limit for the \hdo\ column density has been derived for the extended 
component. On the basis of this improved fit to the molecular
lines, we favor model $D$ over model $C$.

\subsection{Models for [C II] and [O I]} \label{sec:cii}

\begin{figure}
\epsscale{1}
\plotone{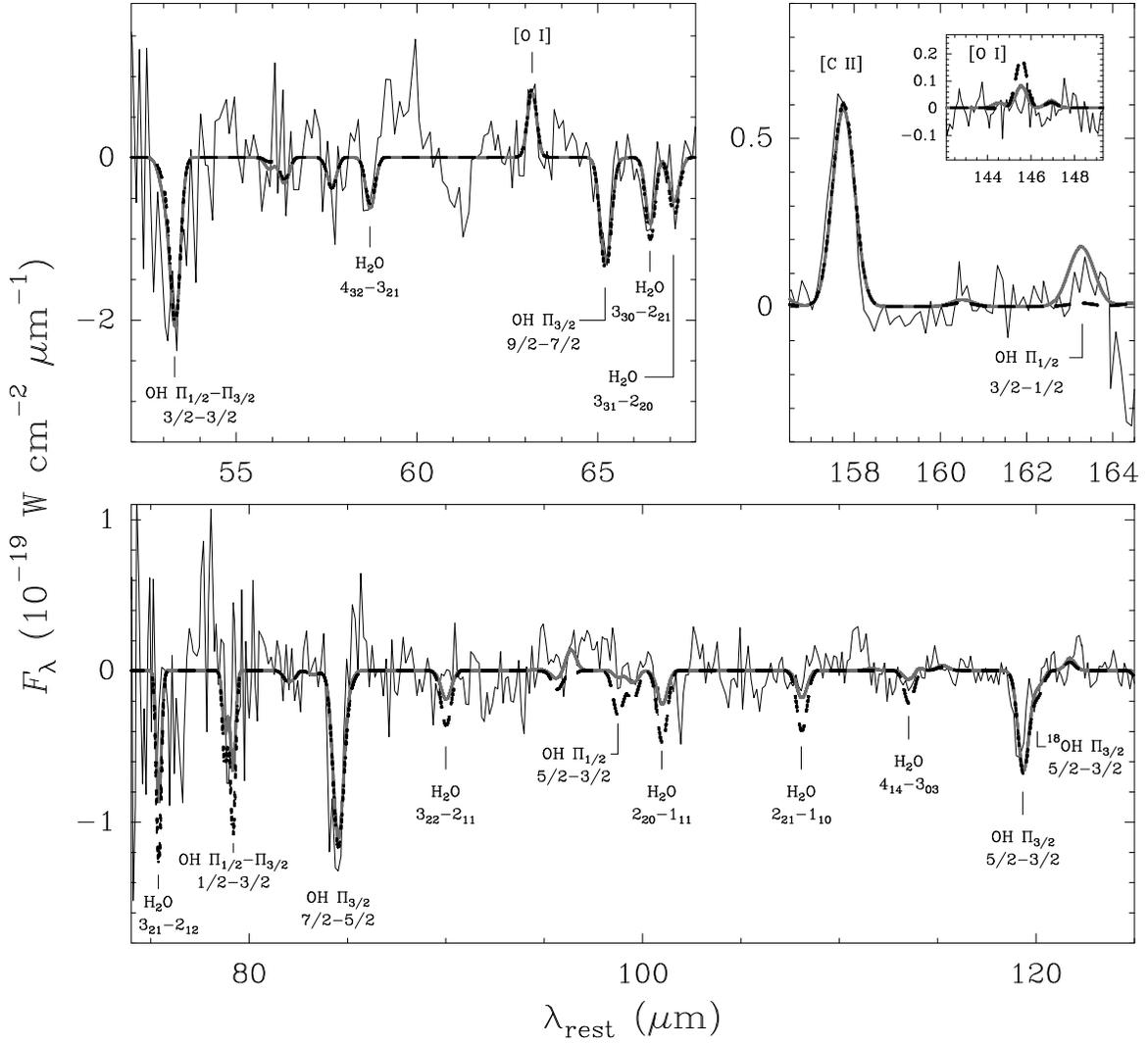}
\caption{Continuum-subtracted spectrum of Mkn 231 compared with two
models for the OH, \hdo, [C II], and [O I] lines. The models are the
single-component $C$ (black-dashed line), and the composite model $D$
(grey-solid line) whose parameters are listed in Tables~\ref{tab:cont} 
and \ref{tab:model}, and
whose predicted continuum SEDs are shown in Fig.~\ref{fig:cont}c-d. The
lines that most contribute to the modeled spectra are labeled.
\label{fig:model}}
\end{figure}

We now check if our preferred models for OH and \hdo\ are consistent 
with the observed [C II] 158 $\mu$m and [O I] 63 $\mu$m lines, 
and in particular if the version with OH and dust coexistent (the mixed 
case), which predicts high OH column densities (Table~\ref{tab:model}), 
is compatible with the low [C II] to FIR flux ratio seen and
previously reported in Mkn 231 \citep{luh03}. In Paper I we
assumed a C$^+$ to OH abundance ratio of 100 (based on standard gas-phase 
carbon abundance, and the abundance of OH 
found in the galactic center and Sgr B2), and found a remarkable agreement
between the predicted and observed [C II] line flux in Arp 220. 
In the present case we have
searched for the required C$^+$ and O$^0$ column densities in models $C$ and
$D$ (mixed case) that fit the observed [C II] and [O I] line fluxes, with 
the additional constraint that the [O I] 145 $\mu$m line is not detected 
to a limit of $4\times10^{-21}$ W cm$^{-2}$ \citep[see also][]{luh03}.
In model $D$, the C$^+$ abundance was assumed to be the same in both 
components. Both the [C II] and [O I] lines are expected to form within
$A_V\sim3$ mag of the surfaces of PDRs, and so we allowed the gas 
temperature to vary between 250 K and 700 K \citep[e.g.,][]{kau99}; 
the results for the [C II] line are not sensitive to $T_k$ as long as $T_k$
is sufficiently higher than the energy of the upper level,
$\approx90$ K. In the regions directly exposed to the FUV radiation field, 
all of the carbon and oxygen are expected to be in their ionized and atomic 
forms, respectively. 
Therefore an O$^0$ to C$^+$ abundance ratio of 2.1, corresponding
to O$^0$ and C$^+$ gas-phase abundances of $3\times10^{-4}$ and 
$1.4\times10^{-4}$ \citep{ss96}, is imposed. Using the collisional rates by
\cite{lau77a,lau77b} for carbon and oxygen, we then searched 
for the density $n({\rm H})$ and column densities $N^{{\rm mix}}({\rm C^+})$ 
and $N^{{\rm mix}}({\rm O^0})$ required to match the line strengths. 
The results for both models $C$ and $D$ are also in Fig.~\ref{fig:model}, 
and the calculated column densities are listed in Table~\ref{tab:model}.

We could not find any satisfactory fit to the three atomic lines with 
model $C$. Since
the continuum is optically thick at $\lambda<150$ $\mu$m, extinction effects 
are important in the [O I] 63 $\mu$m line, and densities higher than
$5\times10^3$ \cmt\ were required to account for it. On the other hand, high
column densities were also needed to fit the [C II] line, and the final result 
is that all combinations that fitted both the [C II] and the [O I] 63 
$\mu$m lines yielded more flux than the upper limit for the undetected [O I] 
145 $\mu$m line. 
The model in Fig.~\ref{fig:model} shows the result 
for $n({\rm H})=8\times10^3$ \cmt\ and $T_k=400$ K. Only high densities 
($\sim10^5$ \cmt) and relatively low temperatures ($\lesssim150$ K) can
approximately match both O$^0$ lines, but then the [C II] line is
underestimated by more than a factor of 2. 

In contrast, model $D$ allowed us to find a more 
satisfactory fit to the three 
lines. As we argue below, the compact-warm component is expected to
give negligible emission in the C$^+$ and O$^0$ lines. The 
extended component $D_{{\rm cold}}$, with its moderate dust temperature 
($T_d=47$ K) and continuum opacity, can explain the [O I] 63 $\mu$m line with 
a density and column density lower than in model $C$, thereby predicting 
a weak 145 $\mu$m line. The model in Fig.~\ref{fig:model} uses for 
$D_{{\rm cold}}$ $n({\rm H})=5\times10^3$ \cmt\ and $T_k=400$ K. 
It is remarkable that when 
we generated the same model but excluded the effect of dust, we
found that the [O I] 63 $\mu$m line was stronger by a factor of 4.2. 
The dust has a small effect on the other two [O I] 145 $\mu$m and 
[C II] 158 $\mu$m lines, but can lower substantially the 
[O I] 63 $\mu$m line because the absorption of 63 $\mu$m 
line-emitted photons by dust grains (i.e., extinction) is more important 
than the additional pumping of O$^0$ atoms by 63 $\mu$m dust-emitted photons. 
Therefore,  the ratio of the [C II] 158 $\mu$m line to the [O I] 63 
$\mu$m line, observed of order unity, is reproduced with higher
densities and temperatures than in previous studies: Fig.~4 
of \cite{kau99} shows that
$I({\rm [O I] \,63 \mu m})/I({\rm [C II] \,158 \mu m})\approx1$ 
for $n({\rm H})=5\times10^3$ \cmt\ and $G_0\sim10^2$, corresponding to a
temperature somewhat higher than 100 K \citep[Fig. 1 of][]{kau99}. 
Including dust leads to a
similar intensity ratio for $T_k=400$ K, or $G_0\gtrsim10^3$. 
On the other hand, our model for $D_{{\rm cold}}$ yields
$I({\rm [C II] \,158 \mu m})/I({\rm [O I] \,145 \mu m})\approx7.5$, which
implies for the above density $G_0\sim10^3$ \citep[see Fig. 6 of][]{luh03}. 
These results suggest that the effect of extinction cannot be neglected when
modeling the [O I] 63 $\mu$m line in sources that are optically thick
at wavelengths shorter than 100 $\mu$m. Self-absorption by foreground
low-excitation clouds may also lower the flux of the [O I] 63 $\mu$m line and 
and to a lesser extent that of the [C II] line \citep[][Paper
I]{fis99,vas02,luh03}.

To compare our results for line intensity ratios involving 
the [O I] 63 $\mu$m line, with the corresponding plots
shown by \cite{luh03}, one should use the versions of our models that ignore
dust, because the intensities given in the models by \cite{kau99} are 
uncorrected for extinction. In our models that ignore 
dust, ${\rm [C\, II]/[O\,I]}\sim0.4$ and 
${\rm ([C\, II]+[O\,I])/I_{FIR}}\sim2\times10^{-3}$. Here, the FIR flux
corresponds to that of component $D_{{\rm cold}}$ (Table~\ref{tab:cont}). 
Using these values in Fig. 4 of \cite{luh03}, one obtains 
$n({\rm H})={\rm a\,few}\times10^3$ \cmt\ 
and $G_0={\rm a\,few}\times10^3$, roughly consistent with our input model
parameters.

The C$^+$ and O$^0$ column densities required to fit the lines in 
model $D_{{\rm cold}}$ are rather high. 
The averaged C$^+$ and O$^0$ abundances 
are $2.2\times10^{-5}$ and $4.6\times10^{-5}$ relative to H nuclei,
respectively. If the C$^+$ to OH abundance ratio is 100 (Paper I), 
the mixed column density of OH is overestimated by a factor of 3. 
In any case, the $N^{{\rm mix}}({\rm C^+})$ values 
indicate that an important fraction of C nuclei is in ionized form, 
$\gtrsim15$\% in model $D_{{\rm cold}}$ for a gas phase carbon abundance of 
$1.4\times10^{-4}$ \citep{ss96}. The corresponding 
mass of atomic gas is $8\times10^8$ \Msun, at least one order of magnitude
higher than the amount estimated in M82 and NGC 278 \citep{wol90,kau99}. 
As mentioned above, C$^+$ is expected to be abundant only in regions 
within $A_V\lesssim2$ mag from the surfaces of PDRs, and the relatively 
high fraction of C$^+$ derived above indicates 
that those regions must be widespread in the Mkn 231 disk. On the other 
hand, one possible interpretation of the high H$_2$ column densities 
required to generate the radiation 
field needed to account for the OH lines, is that we are viewing
a number of PDRs that overlap along the line of sight
(see also \S\ref{sec:discussion}).

We now check the possibility that a significant fraction of the [C II] 
158 $\mu$m emission arises from a more extended region, like the outer 
disk found by DS98 in CO.
For an outer disk mass of $2\times10^9$  \Msun\ and density of $10^2$ cm$^{-3}$
(DS98), and using an upper limit for the C$^+$ abundance of 
$\leq10^{-4}$ and collisional rates at 250 K \citep[from][]{lau77a}, 
we calculate that $\lesssim1/4$ 
of the observed [C II] line arises from this extended region. 
Therefore we do not expect much contribution from it, unless the C$^+$ is 
concentrated in clumps of much higher density.

Our model results can be used to re-examine the issue of the 
[C II] line deficit found in ULIRGs. 
Contrary to what might be expected from the low [C II] to FIR
luminosity ratio in Mkn 231, relatively high average abundances of 
C$^+$ are required in $D_{{\rm cold}}$ 
to reproduce the [C II] line. Explanations based
on self-absorption effects, extinction by dust, effects of dust-bounded HII
regions, high $G_0/n$, and high-formation rate of CO at the expense 
of C$^+$ have been invoked to account for the [C II] deficit
\citep[][Paper I, and references therein]{luh98,luh03,mal01,pap07}. 
In our models for Mkn 231, the high abundance of C$^+$  
yields the observed low [C II] to FIR luminosity ratio
because of the high luminosity-to-gas-mass ratio ($L/M$)
of our continuum models. In both models $C$ and $D$, 
an overall $L/M\approx500$ \Lsun/\Msun\ is
derived (Table~\ref{tab:cont}), where the gas mass includes both the
atomic and molecular components. We argue that, if this extreme $L/M$ ratio 
applies to the whole region from which the FIR
emission arises, as appears to be 
the case for Mkn 231, the [C II] line strength relative to 
the infrared emission can never attain values in excess of 0.1\%. 
The [C II] luminosity emitted per unit of total gas mass is given by
\begin{equation}
\phi_{{\rm [C II]}}=\frac{1}{1+\frac{g_l}{g_u}\exp\{E_u/kT_{ex}\}}
\times n_{{\rm C^+}} A_{ul} h\nu,
\label{eq:cii}
\end{equation}
where $g_l$ and $g_u$ are the degeneracies of the lower and upper level,
$E_u$ is the upper level energy, $A_{ul}$ is the Einstein coefficient for
spontaneous emission, and $n_{{\rm C^+}}$ denotes the number of C$^+$ ions
per unit of total gas mass (including all components). 
It is assumed in eq.~(\ref{eq:cii}) that the line
is optically thin. In the limit $T_{ex}\gg E_u/k$ one obtains
\begin{equation}
\phi_{{\rm [C II]}}\approx 0.5 \times \frac{\chi({\rm C^+})}{10^{-4}} \,\,
{\rm \frac{L_{\odot}}{M_{\odot}}, }
\label{eq:ciib}
\end{equation}
where $\chi({\rm C^+})$ denotes the average C$^+$ abundance relative to total
H nuclei. Therefore, the expected [C II]-to-IR emission is 
\begin{equation}
\frac{\phi_{{\rm [C II]}}}{L/M}\approx 10^{-3} \times
\frac{\chi({\rm C^+})}{10^{-4}} 
\label{eq:ciic}
\end{equation}
for $L/M=500$ \Lsun/\Msun\footnote{For simplicity, the luminosity
accounts here for the total infrared emission as derived in our models,
whereas in previous studies the FIR emission was defined between 40
and 120 $\mu$m; the correction factor is less than 2 \citep[see][]{san96}.}.
Hence, even assuming the highest nearly possible 
$\chi({\rm C^+})\sim10^{-4}$, the [C II]-to-IR ratio is not expected to be 
above $10^{-3}$ for such high values of $L/M$. The reason is simply the 
limited reservoir of gas, and hence of C$^+$ ions, per unit of output 
power. High $L/M$ values are in Mkn 231 primarily due to the contribution of
the AGN, but could also be associated with high star formation efficiencies
\citep[maybe extreme, see][]{sco04}: 
if we attribute the total infrared luminosity to FUV heated grains in 
PDRs and use the units of $G_0$ (i.e. the Habing Field of
$1.6\times10^{-3}$ ergs cm$^{-2}$ s$^{-1}$) to calculate the $L/M$ for a 
PDR column density of $A_v = 10$ mag with the standard 
gas-to-extinction conversion $N({\rm H+2H_2})/A_v \sim 2.5 \times 10^{21}$ 
cm$^{-2}$ mag$^{-1}$ \citep[e.g.,][]{dic78}, we find that the value of $G_0$ 
that corresponds to $L/M=500$ \Lsun/\Msun\ is $2.5\times10^4$, 
which we express for the general case as
\begin{equation}
\frac{L}{M}= 500 \times \left(\frac{G_0}{2.5\times10^4}\right) 
\times \left(\frac{10\,\,{\rm mag}}{A_v}\right)
\,\, {\rm \frac{L_\odot}{M_\odot}},
\label{eq:lm}
\end{equation}
so that high $L/M$ values would indicate high $G_0$ values.
For PDRs with $G_0\sim2.5\times10^4$ and $A_v\sim10$ mag,
the average C$^+$ abundance is $\sim2\times10^{-5}$ (i.e., carbon is 
ionized within $\sim2$ mag of the surfaces of PDRs) and 
$\phi_{{\rm [C II]}}/(L/M)$ will attain the lowest values observed 
in ULIRGs. 

In our favored model $D$, the component 
$D_{{\rm warm}}$ ($L/M=3270$ \Lsun/\Msun), which emits most of the IR 
luminosity, could only account for the [C II] line if 
{\em essentially all} gas-phase carbon 
were in ionized form. However, this seems implausible, 
as the C$^+$ abundance in XDRs is $\chi({\rm C^+})\sim {\rm a \,\, few \times}
10^{-5}$  \citep{mei05}, and FUV photons from stars are absorbed
in thin shells of dust on the surfaces of PDRs. With the assumed column
densities in Table~\ref{tab:model}, the calculated contribution from 
$D_{{\rm warm}}$ to the observed [C II] emission is only 8\%, 
the [O I] 145 $\mu$m flux is 15\% of the upper 
limit, and the [O I] 63 $\mu$m is predicted in weak {\it absorption}
against the continuum. (In this calculation, $T_k=10^3$ K and 
$n=2\times10^3$ \cmt\ are assumed, so that the atomic region is in pressure
equilibrium with the cooler molecular gas with $T_k\sim10^2$ K and 
$n=2\times10^4$ \cmt.)
It is therefore expected that the strongest source of infrared radiation 
(the warm component) gives low [C II] and [O I] line emission even if one
assumes relatively high averaged fractional abundances for C$^+$ and O$^0$. 
As in \cite{luh03}, we then invoke a component that emits strongly in the 
FIR but yields weak emission in the [C II] and [O I] lines. 
In contrast to \cite{luh03}, however, we leave open 
the possibility that $D_{{\rm warm}}$ has a relatively high 
C$^+$ abundance, as expected from both XDR and PDR chemistry.
In starburst and normal galaxies, the moderate star formation efficiencies 
\citep*[e.g., $\sim15$ \Lsun/\Msun\ in NGC 1068;][]{pla91}, implying lower
values of $G_0$, will not place such a strong limit on the C$^+$ cooling 
rate relative to the infrared emission.
We suggest, then, that the [C II] 158 $\mu$m line deficit found in ULIRGs
may be associated with a high $L/M$ ratio that, allowing 
a relatively high C$^+$ abundance, indicates an important contribution to
the FIR emission by an AGN, and/or a high star-formation efficiency.

\section{Discussion and conclusions} \label{sec:discussion}

Simultaneous modeling of the FIR continuum and line absorptions is 
the only consistent way to derive from FIR observations the physical 
and chemical properties of bright-infrared galaxies, where due to 
high radiation densities, the OH and \hdo\ lines are pumped through 
absorption of continuum photons. 
In Mkn 231, the high-lying 65-68 $\mu$m OH and \hdo\ lines indicate 
the presence of a compact ($\sim100$ pc) and warm ($\sim100$ K) region, 
presumably located around the AGN, with high continuum opacity 
($\tau_{{\rm 100\mu m}}\sim1$). The size of this region is comparable to
(though somewhat larger than) the inner nuclear disk where the OH 
mega-masers have been observed from VLBI observations \citep{klo03}, 
and is similar to the spatial extent of the regions where the bulk 
of the OH mega-maser emission is generated \citep[regions
C, NE, and SW in Fig. 5 of][]{ric05}. This strongly suggests 
that we are observing the FIR counterpart of the 18 cm OH 
emission \citep[e.g.,][]{ski97}. 
\cite{ric05} estimate $N({\rm OH})\sim1.5\times10^{15}$ \cmd\ in 1 pc clouds
where $\Delta V\approx1.7$ \kms; therefore, $N({\rm OH})/\Delta V \sim
10^{15}$ \cmd/(\kms). In both the warm and cold components of model $D$, we
have obtained 
$N^{{\rm scr}}({\rm OH})/\Delta V \sim 1.5\times 10^{15}$ \cmd/(\kms), in rough
agreement with the estimates by \cite{ric05}. The ``screen'' case is  
most appropriate for comparison here because the OH megamasers amplify the
background radio continuum emission.

The nuclear region is surrounded by a more extended ($R_{eq}\sim350$ pc) 
and colder ($\sim45$ K) region, which produces the bulk of the [C II] and 
[O I] line emissions and contributes to the low-lying OH lines. Allowing for
an area filling factor of 0.6-0.7, the total extent 
of the FIR emission is then comparable to the size of the CO (inner) disk 
\citep[][DS98]{bry96}. The inferred densities ($n({\rm H_2})\sim10^4$ \cmt) 
may explain the submillimeter CO emission detected by \cite{pap07}, and 
somewhat lower densities (${\rm a \,\, few}\times10^3$ \cmt) can account for 
the [O I] line emission. We do not however rule out the possibility that
some of the FIR emission arises from a more extended region, as the outermost
disk observed by DS98 in the millimeter CO lines or the much more extended
region observed in soft X-rays \citep{gal02}.

The FIR emission from $D_{{\rm warm}}$ is probably dominated by 
the AGN, as pointed out in \S\ref{sec:continuum}. \cite{dav04} estimated
that, within a radius of 330 pc, the starburst luminosity accounts for
$(0.8-1.3)\times10^{12}$ \Lsun; since the luminosity for the component
$D_{{\rm cold}}$ is $\approx7.5\times10^{11}$ \Lsun, the 
starburst luminosity contribution from $D_{{\rm warm}}$ is most probably 
$\lesssim6\times10^{11}$ \Lsun. \cite{dav04} also
estimated that, within 100 pc of the AGN (the size of $D_{{\rm warm}}$), the 
starburst contribution is $(3.2-4.7)\times10^{11}$ \Lsun, in agreement with 
the afore mentioned upper limit. On the other hand, the extended 
($360\times260$ pc$^2$) radio continuum 1.4 GHz emission detected by 
\cite{car98} is compatible with both the whole FIR emission arising
from the starburst \citep[yielding an IR-to-radio flux density ratio of
$Q\sim2.5$,][]{car98}, and only the $D_{{\rm cold}}$ FIR emission 
arising the starburst (yielding $Q=2.1$). If one now assumes that the 
starburst contribution to the FIR emission of $D_{{\rm warm}}$  
is $3\times10^{11}$ \Lsun, 
the resulting $Q$ is 2.3 \citep[in agreement with the median value of
$Q=2.3\pm0.2$ in normal galaxies;][and references therein]{con92}. Also, the
surface brightness and luminosity-to-mass ratio of $D_{{\rm warm}}$ 
due to only the starburst component would then be 
$2\times10^{12}$ \Lsun\ kpc$^{-2}$ and 550 
\Lsun/\Msun, which are probably upper limits for a starburst. 
In summary, our preferred model $D$ favors that $\sim2/3$ of the bolometric 
luminosity from Mkn 231 is due to the AGN.

Although the chemistry of the extended starburst $D_{{\rm cold}}$
component, where the lower-lying OH lines and the bulk of the [C II] 158
$\mu$m and [O I] 63 $\mu$m lines are generated,
appears to be dominated by PDRs, the dominant chemistry of the warm component
may be of a different nature. Since the compactness of $D_{{\rm warm}}$
suggests that its FIR emission is dominated by the AGN, and given the high
X-ray intrinsic luminosity of the AGN ($\sim10^{44}$ erg s$^{-1}$) in the 2-10
keV band \citep{bra04}, we consider the possibility
that the derived OH and \hdo\ column densities can be better explained in
terms of XDR chemistry. Calculations by \cite{mei05} show that,
in dense XDRs with $N_{{\rm H}}\sim10^{24}$ cm$^{-2}$, the OH and \hdo\ column
densities are expected to attain values of $\sim10^{18}$ cm$^{-2}$, similar to
the values that we have derived in the mixed case. Furthermore, the
Compton-thick screen with $N_{{\rm H}}\sim2\times10^{24}$ cm$^{-2}$ required
to block the primary X-ray emission \citep{bra04} could be partially
identified with $D_{{\rm warm}}$, with similar column density. 
Therefore an XDR may naturally explain the observations of
high-lying OH and \hdo\ lines in Mkn 231. Nevertheless, both OH and \hdo\ are
also tracers of starburst chemistry, and a number of PDRs
and hot cores overlapping along the line of sight could in principle yield the
column densities required to explain our observations at least in the screen
case, and could then mimic an XDR. We conclude that further observations are
still required to discern between XDR and starburst chemistry.

We suggest  
that the [C II] 158 $\mu$m line deficit in Mkn 231, and also
probably in Arp 220 (see Paper I), is primarily due to a component that 
dominates the FIR emission but emits weakly in the [C II] line, 
as previously suggested by \cite{luh03}. However, we also suggest that
this component may still be rich in C$^+$, but with an extreme 
luminosity-to-gas-mass ratio that limits the [C II] luminosity per unit of 
luminous power in the continuum. Our derived luminosity to H$_2$ mass 
ratio of 500 \Lsun/\Msun\ is high; future studies will indicate 
whether it can be applied to other ULIRGs with similar [C II] deficits.

Given the high C$^+$ column densities (a few $\times10^{19}$ \cmd) required to
account for the [C II] 158 $\mu$m line emission, the H$_2$ columns derived
toward the extended component are unlikely to be associated with 
non-overlapping star-forming regions surrounded by optically thick 
envelopes. A single PDR is not expected to have $N({\rm C^+})$ in excess 
of ${\rm a \,\,few}\times10^{17}$ \cmd. A scenario with a crowded 
population of PDRs overlapping along the line of sight is 
therefore more plausible. 
If a typical {\it single} OB stellar cluster emits up to 
$10^{39}$ erg s$^{-1}$, the maximum value found by \cite{sco01} 
in M51 and also the luminosity of W49, then the number of single clusters to
produce the Mkn 231 starburst luminosity of $10^{12}$ \Lsun\ is 
$\gtrsim4\times10^6$. If these
clusters are concentrated in a disk of radius 460 pc and thickness 25 pc 
(DS98), the mean distance between neighboor clusters is only $\lesssim1.5$
pc \citep[see also][]{ket92}. Merging of PDRs may be compatible with 
some area filling factor if, for example, some spiral 
structure is invoked within the disk.

In NGC 1068, \cite{spi05} observed the OH 119 $\mu$m line in emission 
against the continuum. On the basis of a possible XDR chemistry, and given 
the corresponding mass, density and temperature that 
characterize the nuclear region, \cite{spi05} suggested that the OH 119 $\mu$m 
emission line could be formed in that region. 
This option also relied on the fact that relatively 
weak FIR emission is expected to arise from the circumnuclear 
disk of the Seyfert 2 galaxy, so that the OH 
119 $\mu$m line can be excited through collisions and emit above the
continuum in these dense and warm environments. The situation
is different for Mkn 231 and Arp 220. In these ULIRGs, the extremely high 
luminosity arising from the nuclear region, together with the high
concentrations of gas there, make the nuclear FIR 
emission component very bright. Any possible emission 
in the 119 $\mu$m line will be obscured by the strong absorption, 
which in these objects is $\sim30$ times stronger than the emission 
feature in NGC 1068. Also, the strong FIR 
radiation density pumps higher-lying OH and \hdo\ levels, thus producing
absorptions in the higher-excitation lines. The OH and \hdo\ molecules are,
therefore, potentially powerful tracers of circumnuclear regions around 
AGNs. Future Herschel observations of Mkn 231 and other sources will 
allow us to apply the models developed in this paper to these sources, 
refine them accordingly, and will certainly 
give new insights into the physical and chemical conditions of 
bright infrared galaxies.

\acknowledgments
E. G-A thanks the Harvard-Smithsonian Center for Astrophysics for its 
hospitality. The authors would like to thank M. Wolfire for helpful
discussions. Research was supported in part by NASA grant NAG5-10659 and 
NASA grant NNX07AH49G. Basic research in infrared astronomy at the Naval 
Research Laboratory is supported by 6.1 base funding.
This research has made use of NASA's Astrophysics Data System.

\clearpage

% Table 1

\begin{deluxetable}{cccccc}
%\tabletypesize{\scriptsize}
%\rotate
\tablecaption{Line fluxes, continuum flux densities, and equivalent 
widths for the lines detected in the ISO/LWS spectrum of Mkn 231 
\label{tab:flux}}
\tablewidth{0pt}
\tablehead{
\colhead{Species} & \colhead{Transition} & 
\colhead{$\lambda_{{\rm rest}}$\tablenotemark{a}} &
\colhead{Line flux\tablenotemark{b}} & \colhead{Continuum\tablenotemark{c}}  
& \colhead{$W$\tablenotemark{d}} \\
& & ($\mu$m) & ($10^{-20}$ W cm$^{-2}$) & 
($10^{-19}$ W cm$^{-2}$ $\mu$m$^{-1}$) & ($10^{-2}$ $\mu$m) }
\startdata
OH & $\Pi_{1/2}-\Pi_{3/2}\,\,3/2-3/2$ & 53.3 & $-7.0\pm2.0$ & 33.5 & $2.1\pm0.6$ \\
${\rm [O\,I]}$ & $^3P_1-^3P_2$ & 63.2 & $2.6\pm0.8$ & 25.0 & $-1.0\pm0.3$ \\
OH & $\Pi_{3/2}\,\,9/2-7/2$ & 65.2 & $-4.1\pm0.6$ & 23.8 & $1.7\pm0.3$ \\
\hdo\ & $3_{30}-2_{21}$ & 66.4 & $-3.3\pm0.9$ & 23.1 & $1.4\pm0.4$ \\
\hdo\ & $3_{31}-2_{20}$ & 67.1 & $-2.2\pm0.9$ & 22.7 & $1.0\pm0.4$ \\
OH & $\Pi_{3/2}\,\,7/2-5/2$ & 84.5 & $-8.6\pm0.7$ & 13.8 & $6.2\pm0.5$ \\
\hdo\ & $2_{20}-1_{11}$ & 101 & $-1.3\pm0.3$ & 8.70 & $1.5\pm0.4$ \\
OH & $\Pi_{3/2}\,\,5/2-3/2$ & 119.3 & $-3.6\pm0.3$ & 5.22 & $6.9\pm0.6$ \\
${\rm [N\,II]}$ & $^3P_{2}-^3P_{1}$ & 121.8 & $1.5\pm0.2$ & 4.72 & $-3.2\pm0.4$ \\
${\rm [C\,II]}$ & $^2P_{3/2}-^2P_{1/2}$ & 157.7 & $3.7\pm0.1$ & 1.73 & $-21.4\pm0.6$ \\
\enddata
\tablenotetext{a}{For OH doublets, an average for the two components is given.}
\tablenotetext{b}{Errors do not include calibration uncertainties in the
  continuum level. Negative (positive) values indicate that the line is
  detected in absorption (emission).}
\tablenotetext{c}{Uncertainties in the continuum level are less than 30\%.}
\tablenotetext{d}{Equivalent widths are positive (negative) for lines observed 
in absorption (emission).}
\end{deluxetable}

% Table 1

\begin{deluxetable}{cccccccccc}
%\tabletypesize{\scriptsize}
%\rotate
\tablecaption{Models for the continuum emission \label{tab:cont}}
\tablewidth{0pt}
\tablehead{
\colhead{Model} & \colhead{Component}   & \colhead{$\tau_{{\rm 100\mu m}}$}   &
\colhead{$R_{eq}$} & \colhead{$N({\rm H}_2)$\tablenotemark{a}}  
& \colhead{$T_{{\rm d}}$} & \colhead{$T_{{\rm rad}}$} & \colhead{$M$} & 
\colhead{$L$}  & \colhead{$L/L_{{\rm IR}}$} \\
   & & & \colhead{(pc)} & \colhead{(\cmd)} & \colhead{(K)} & \colhead{(K)} &
\colhead{(\Msun)} & \colhead{(\Lsun)} & }
\startdata
$A$ & - & $6.8\times10^{-2}$ & 1830 & $4.6\times10^{22}$   & variable\tablenotemark{b} 
& 23 & $1.0\times10^{10}$ & $2.4\times10^{12}$ & 0.72 \\
$B$ & - & $0.55$ & 400 & $3.7\times10^{23}$  & 55 & 44 & $4.0\times10^{9}$ 
& $1.7\times10^{12}$ & 0.50 \\
$C$ & - & $2.2$ & 200 & $1.5\times10^{24}$  & 74 & 71 & $4.0\times10^{9}$ 
& $2.0\times10^{12}$ & 0.61 \\
$D$ & warm & 1.1 & 106 & $7.2\times10^{23}$  & 100 & 85 & $5.5\times10^{8}$ 
& $1.8\times10^{12}$ & 0.56 \\
$D$ & cold & 0.97 & 350 & $6.5\times10^{23}$ & 47 & 42 & $5.3\times10^{9}$ 
& $7.4\times10^{11}$ & 0.22 \\
\enddata
\tablenotetext{a}{A gas-to-dust mass ratio of 100 is assumed, together with a
  mass-absorption coefficient for dust of 44 cm$^2$ g$^{-1}$ at 100 $\mu$m.}
\tablenotetext{b}{The temperature is calculated from the 
balance between heating and cooling; see text for details.}
\end{deluxetable}

\clearpage

% Table 2

\begin{deluxetable}{lccc}
%\tabletypesize{\scriptsize}
%\rotate
\tablecaption{Models for molecules and atoms \label{tab:model}}
\tablewidth{0pt}
\tablehead{
\colhead{Model\tablenotemark{a}$\rightarrow$} & \colhead{$C$} & 
\colhead{$D$} & \colhead{$D$} \\
 & & \colhead{(warm)} & \colhead{(cold)} }
\startdata
$N({\rm H_2})$ (\cmt) & $1.5\times10^{24}$ & $7.2\times10^{23}$ & $6.5\times10^{23}$   \\
$\Delta V$ (km s$^{-1}$) & 40  & 60  & 40 \\
$N^{{\rm scr}}({\rm OH})$\tablenotemark{b} (\cmd) & $1.0\times10^{17}$ & $9.2\times10^{16}$  & $6.1\times10^{16}$  \\
$N^{{\rm mix}}({\rm OH})$\tablenotemark{b} (\cmd) & $4.5\times10^{18}$ & $1.4\times10^{18}$  & $9.7\times10^{17}$  \\
$\chi$(OH)\tablenotemark{c}                          & $3.0\times10^{-6}$ & $2.0\times10^{-6}$  & $1.5\times10^{-6}$  \\
$N^{{\rm scr}}({\rm H_2O})$ (\cmd) & $5.4\times10^{16}$ & $6.1\times10^{16}$  & $<5.0\times10^{15}$  \\
$N^{{\rm mix}}({\rm H_2O})$ (\cmd) & $2.0\times10^{18}$ & $9.6\times10^{17}$ & $<1.0\times10^{17}$ 
\\
$\chi$(\hdo)\tablenotemark{c}        & $1.3\times10^{-6}$ & $1.3\times10^{-6}$  & $<1.5\times10^{-7}$ \\
$N^{{\rm mix}}({\rm C^+})$\tablenotemark{d} (\cmd) & $2.3\times10^{20}$ & $3.1\times10^{19}$\tablenotemark{e} & $2.8\times10^{19}$\tablenotemark{e}   \\
$N^{{\rm mix}}({\rm O})$\tablenotemark{d} (\cmd)  & $4.9\times10^{20}$ & $6.7\times10^{19}$\tablenotemark{e}  &  $6.0\times10^{19}$\tablenotemark{e}  \\
\enddata
\tablenotetext{a}{Models correspond to the continuum models listed in 
Table~\ref{tab:cont} and shown in Fig.~\ref{fig:cont}c and d.}
\tablenotetext{b}{$N^{{\rm scr}}(X)$ is the column density of a shell of
species $X$ surrounding the continuum source, and $N^{{\rm mix}}(X)$ is the
column density for the case that species $X$ is evenly mixed with the dust.}
\tablenotetext{c}{Abundances are given relative to H$_2$ using the 
$N^{{\rm mix}}(X)$ values.}
\tablenotetext{d}{The column densities of C$^+$ and O are computed by assuming
  gas at 400 K, and a O$^0$ to C$^+$ abundance ratio of 2.1.}
\tablenotetext{e}{Equal averaged abundances are assumed in the warm and cold
components of model $D$.}
\end{deluxetable}

\clearpage

%Figure 1

%Figure 2

%Figure 3

%Figure 4

%Figure 5

%Figure 6

\end{document}